# Electrostatic contribution to DNA condensation – application of 'energy minimization' in a simple model in strong Coulomb coupling regime.

**Arup K. Mukherjee**

Department of Physics, Chancellor College, Box 280, Zomba,
University of Malawi. Malawi.
E-mail: akmukherjee11@hotmail.com

## Abstract

Bending of DNA from a straight rod to a circular form in presence of any of the mono-, di-, tri- or tetravalent counterions has been simulated in strong Coulomb coupling environment employing a previously developed energy minimization simulation technique. The inherent characteristics of the simulation technique allow monitoring the required electrostatic contribution to the bending. The curvature of the bending has been found to play crucial roles in facilitating electrostatic attractive potential energy. The total electrostatic potential energy has been found to decrease with bending which indicates that bending a straight DNA to a circular form or to a toroidal form in presence of neutralizing counterions is energetically favorable and practically is a spontaneous phenomenon.

# I. Introduction

For the past four decades DNA condensation has drawn much attention [1-8] because it is a popular research theme in a plethora of areas of science like biophysics, biochemistry, molecular biology, polymer physics, condense matter physics and biotechnology. In biotechnology long genomic DNAs are required to be condensate to transfer from test tube to living cells for gene therapy application [6,7,8]. Toroidal structures are more interesting in DNA condensation since direct observation suggested toroidal structures in the lysates of bacteriophages and the generation of DNA toroids in vitro by virus and sperm-cell DNA-condensing agents [9].

Condensation of DNA depends on a number of factors, such as, length of the DNA, type of the solvent (especially the dielectric constant) and condensing agents [8], solute (cations) and its valence in the solution, solution concentration, temperature etc. The required counterion valance also depends on the solvents for condensation. Cherstvy and Kornyshev et al [10, 11] reported that it was not always the valence of the counterions only but the type was also important. For example $Mg^{++}, Ca^{++}$ induce aggregation of double stranded DNA but do not provoke condensation while $Mn^{++}, Cd^{++}$ induce both precipitation and condensation even though all of them are divalent cations. Cation size is another significant factor since bending angle depends on it. Wide verities of reports on proper combinations of condensing agents and solvents and their concentrations for expected DNA condensation are abundant in literature [8 and references therein]. The behavior of DNA (morphology of condensation) under various types of condensing agents and their solvents, required concentrations, solution temperatures etc are yet to be understood clearly. The problem is that when one tries analytically to generalize (or even to develop a model of DNA condensation especially applicable to study bacteriophages and sperm cells [9]) DNA behaviors under various applied conditions one can find it generally difficult due to those apparently erratic responses of DNA. The complexity of inclusion of so many factors related to the environment surrounding the DNA adds another difficulty to understand the real physics behind condensation.

Even though DNA condensation by alcohols and neutral or anionic polymers has been reported [8], there is a general conjecture that the electrostatic repulsions due to highly negatively charged phosphate groups of DNA backbone reduce the flexibility of DNA to condense it easily. Thus it is required to neutralize those for condensation. Quantitative analysis suggests that neutralization of approximately 90% or more of those charges favor condensation [12]. This suggests also that except few cases electrostatic interactions play the crucial roles in DNA condensation. The interplay of electrostatic interactions is not confined to DNA condensation only but to DNA twisting, stretching, helical deformation and loop formation also [13, 14, 15]. Persistence length strongly depends on salt concentration and dielectric core [14]. Twisting and bending persistence lengths are largely influenced by low dielectric core than water permeable core of a DNA [16]. As most of the biological organisms are highly charged, electrostatic interactions govern nearly about all of their biophysical and biochemical activities. A very recent paper by A. G. Cherstvy [17] describes in detail how electrostatics are involved in biological systems including almost all relevant current research activities, their outputs and future aspects.

In this paper, mainly to understand the electrostatic role on DNA condensation apart from any other external influences an attempt has been taken to simulate a simplified picture of DNA condensation under a specific case of strong Coulomb coupling employing a previously developed Energy minimization (EM) technique [20]. In this dynamic picture of simulation, a flexible straight rod-like DNA bends gradually to a circle. For every degree of bending the change in total electrostatic energy has been calculated only considering the electrostatic interactions between macroion and its neutralizing cations attached on the surface of the macroins. In this simplified picture, the total minimized energies (calculated minimizing counterion mutual positions) clearly indicate that the bending of DNA largely depends on its geometrical facility on electrostatic interactions. It has been seen that every degree of bending of a straight DNA towards a circle is increasingly energetically favorable implying a rapid bending to a circle and finally to a tightly wound toroid (or other compact morphology) for stability. From this study one can observe the role of electrostatics alone on DNA bending and strong Couloumb coupling regime is the ideal environment for this.

In a strong Coulomb coupling environment all the counterions in the solution (in this study the solution concentration can be envisaged in such a way that the total counterion charge is the same as the total bare charge of the macroion in a salt free condition. This is also required for MC simulation to check some aspects of this study.) surrounding a charged macroion surface are bound with that surface via only electrostatic interaction. If the counterion concentration surrounding the charged surface can be expressed as $\rho(x) = \rho_o \exp[-Z_C e\Psi/k_B T] \to 0$ as $x \to \infty$, where $\rho_o$ is bulk concentration and $\Psi(r)$ is the self consistent electrostatic potential, then the solution of Poisson-Boltzmann equation assumes Gouy-Chapman form $\rho(x) = \frac{1}{2\pi Z_C^2 l_B} \frac{1}{(\lambda+x)^2}$, where $\lambda = e/2\pi Z_C l_B \sigma$ is the Gouy-Chapman (GC) length, $l_B = e^2/4\pi\varepsilon_o\varepsilon_r k_B T$ is the Bjerrum length, $Z_C$ is the counterion valance, $\sigma$ is the surface charge density and $k_B$ is the Boltzmann constant [18]. $\sigma \cong 0.94$ for a straight cylinder of length 51 nm, radius 1 nm and an axial line charge of -300e (as considered throughout this study). $l_B \cong 0.7$ nm for water solvent. Thus the Gouy-Chapman length $\lambda \cong 0.24/Z_C$ nm, which implies, as the radius of counterions is 0.18 nm, that all the counterions (except monovalent) are attached with the cylinder surface. Also, as $\lambda$ is directly proportional to the solvent relative permittivity $\varepsilon_r$, a lower value of $\varepsilon_r$ (= 16) has been chosen in some literatures [19, 20] to ensure strong Coulomb coupling. The above discussion apparently shows that for the present system pure water ($\varepsilon_r$ = 80) can also be used as solvent. The coupling can be termed as 'high' when $\frac{l_B}{\lambda} \gg 1$ [21] and in case of water solvent $\frac{l_B}{\lambda} = Z_C\left(\frac{0.7}{0.24}\right) = 2.92 Z_C$. However, a series of standard canonical Metropolis Monte Carlo (MC) simulation studies (which have been performed for this work using the above stated system, only to check the validity of the above statement) indicate that water solution at room temperature is not a very perfect condition to condense all multivalent counterions (except tetravalent or higher valance counterions) on the macroion cylinder. According to Manning-Oosawa model [22] a fraction $\varphi$ of the counterions can condense on the macroion surface when the average spacing b between the charges becomes smaller than $l_B$. The fraction of condensed counterions is expressed as $\varphi = 1 - \frac{b}{Zl_B}$, where Z is the condensed counterion valance. As $l_B$ is inversely proportional to temperature and dielectric constant, decrease in either one or both can increase $l_B$ which leads to increase in $\varphi$ for any specific Z. In a recent experimental work [23] Wen and Tang showed that the effects of both the parameters could be combined to a single one and that was $l_B$. This has also been tested and found to be true again employing the MC simulation using the same system. Basically, the WC model is a zero temperature approximation (even though it has been extended later to finite temperature by treating the counterions as strongly correlated liquid [24,25]) in which physiological conditions can hardly be maintained (under the freezing temperature of water). Besides that thermal agitation is required for mobility of counterions (according to the prediction of Oosawa model) to achieve stronger correlations while in WC model thermal agitation destroys the lattice structure of the ionic crystals and thus leads to weaker correlations [23]. Thus for the present study a fixed temperature of 275 K and a dielectric constant of 20 of the solvent (water solution with ethanol and counterions with no added salt) can be considered ($l_B \sim$ 3.04 nm), even though in EM technique [20] all those parameters appear as just pre-factors and do not interfere with the ionic distributions on the macroion surface. Ethanol is required only to decrease the dielectric constant of the solution. This solution condition can also maintain the strong Coulomb coupling environment in liquid water. Note that the MC simulations within the primitive electrolyte model imply uniform continuum dielectrics with the dielectric permittivity of the solvent surrounding the DNA. In fact, the molecules of the solvent around the DNA can be strongly oriented in the electrostatic field of the highly charged DNA which can make the dielectric permittivity much lower than in the bulk solution. According to some estimations [26,27] the dielectric constant of water solvent may be as low as 6 – 30 at very near to the DNA surface. Also one needs to consider the fact that the dielectric constant of the DNA itself is much lower [16] than that of water, nevertheless, the rule of thumb of the most theoretical and computational works accept no dielectric discontinuity between the solution and the macroion to avoid the image charge effect.

In this study overcharging [18-21,29] has not been considered. As stated earlier, the water solution contains only the number of counterions (of any valance and without mixture) which can barely compensate the macroion charge (not overcompensate), there is no opportunity for the macroion to be overcharged.

## II. Model and Simulation Methods

As stated above the system considered in this study is comprised of a cylinder (macroion) of length ($L$) 51 nm and of radius (r) 1 nm (mimicking a c-DNA) with bare macroion charge $Q = -Z_m |e|$ surrounded by a number $N_C$ of small spherical counterions with charge $q = Z_C |e|$ and radius ($R_C$) 0.18 nm so that $Z_m = Z_C N_C = 300$ in the neutral state. $Z_m$ and $Z_C$ are the macroion and counterion valances respectively. The charge of the cylinder has been considered as being comprised of very closely distributed point charges of magnitude $z_i |e|$ in a line along the axis of the cylinder so that $Z_m |e| = \sum_{i=1}^{n} z_i |e|$, where n is the total number of such points. This has been considered instead of a continuous line charge (Manning conception) [30] to facilitate the calculations by avoiding frequent solutions of generally non-complete elliptic integrals. The macroion charge distribution will be further discussed later in the concluding section. To calculate the total minimized electrostatic energy a previously developed [20] energy minimization simulation technique has been employed. The energy minimization technique is a simulation technique that calculates the counterion positions on the surface of the macroion by minimizing the distances among them for which the total electrostatic potential energy of the macroion-counterion system yields very near to the lowest possible (ground state) energy. Here one needs to consider all the counterions are always at a constant counterion-macroion distance ($\tau = r + R_C$) of closest approach. This is an intrinsic requisite of the technique. This condition is also required to maintain the environment of strong Coulomb coupling [19, 20]. As the macroion has been considered as hard core the Lennard-Jones potential calculations are not required for this study. Note that it has been shown earlier [20] that the energy minimization technique produces exactly the same results as MD or MC under strong Coulomb coupling condition. But this technique is rather simple and easy to use (without considering a cell or a solution surrounding the DNA) for any macroion geometry. The technique converses rapidly and thus economic in terms of computer time.

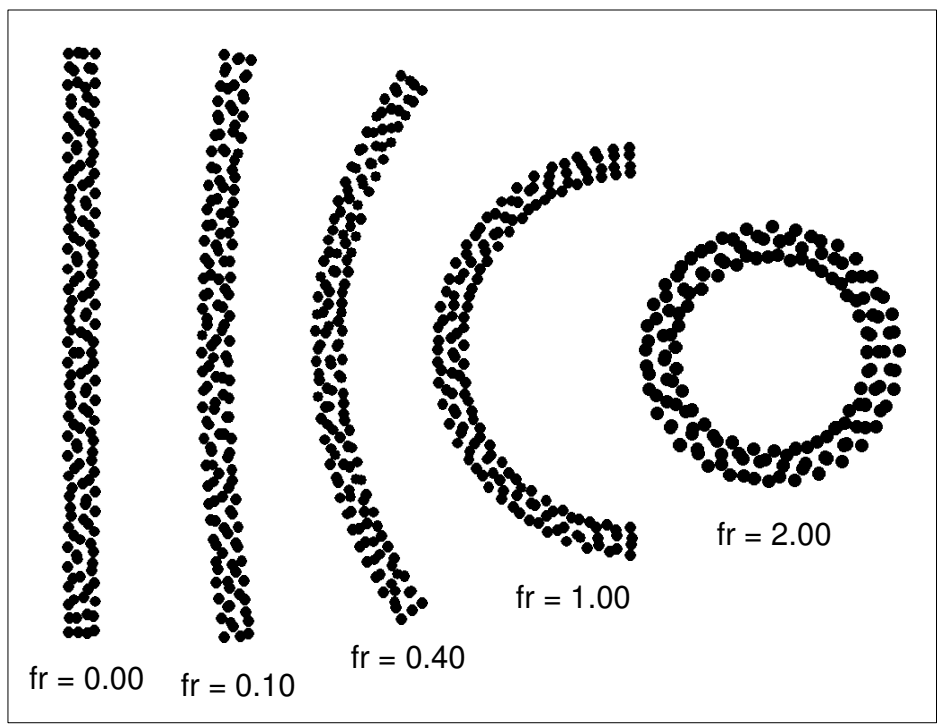

**Figure 1.** The process of gradual bending of a straight cylinder (surrounded by energy minimized counterions of any valance) to a circle. The black dots represent counterions. 'fr' is the bending fraction that varies from 0 (straight line) to 2 (circle).

For the purpose of the present study, a straight cylinder has been considered at first and the total minimized energy (counterion-counterion plus counterion-macroion) has been calculated. Next, the cylinder has been bent a little to an arc of a circle and the minimized energy has been measured by the same way as in the case of the straight cylinder. The process of bending continues until the cylinder forms a circle (which is a pre-step to a toroid). In every step of bending (fr) the total minimized energy has been calculated. The bending fraction 'fr' is defined as $L/\pi R$, where R is the radius of curvature. Obviously 'fr' can take up any value between zero and two. Figure 1 depicts the process of bending.

The total electrostatic energy of the counterion-macroion system reads,

$$E_{Coul} = \frac{|e|^2}{4\pi\varepsilon_o\varepsilon_r}\left[\sum_{i=1}^{n}\sum_{j=1}^{m}\frac{z_j Z_C}{\tau_{ij}} + \sum_{i<j}\frac{Z_C^2}{r_{ij}}\right] \qquad [1]$$

where $\varepsilon_r$ is the relative permittivity, $\tau_{ij}$ is the distance between any counterion i and a point macroion charge j and $r_{ij}$ is the separation between any two counterions i and j. Equation 1 can be written as

$$E_{Coul}/k_B T = l_B\left[\sum_{i=1}^{n}\sum_{j=1}^{m}\frac{z_j Z_C}{\tau_{ij}} + \sum_{i<j}\frac{Z_C^2}{r_{ij}}\right] \qquad [2]$$

It is worth to mention that the relative permittivity of the cylinder mimicking the c-DNA has been considered identical to that of its out side (the surrounding counterions) to avoid image charge problems. The Bjerrum lelgth has been taken as 30.38 Å all over the study. As the simulation technique converges significantly rapidly, around 300,000 moves per counterion is sufficient to reach very close to the lowest possible energy state.

**III. Results and discussion**

The total minimized energy has been seen to decrease when the straight cylinder is gradually bent to a circle. Figure 2 shows the detail energy profile with the increase of bending fraction fr. The Y –axis of figure 2 corresponds to the total electrostatic energy per unit thermal energy. Note that, as stated earlier, the shapes of the curves do not depend on $\varepsilon_r$ and T of solvent. Those are just pre-factors which determine the magnitudes of the calculated energies only (see equation 2). In fact, the study has been performed considering non-screened Coulomb interactions. If screening was considered then the Debye-Huckel potential would act between the charges as [16]

$$\Psi_{DH}(r_{ij}) = \frac{e^2}{4\pi\varepsilon_o\varepsilon_r k_B T} \frac{\exp[-\kappa r_{ij}]}{r_{ij}} = \left(\frac{E_{Coul}}{k_B T}\right)\exp[-kr_{ij}] \qquad [3]$$

where $\kappa^{-1} = \frac{1}{\sqrt{4\pi l_B \rho_o}}$ is the Debye screening length (salt free), $\rho_o$ is the bulk concentration of monovalent counterion species and $r_{ij}$ is the distance between either a counterion and a macroion point charge or a counterion and another counterion. But the decrease in total electrostatic interaction with bending would definitely remain visible. However, due to the applied restrictions on solution concentration (stated in introductory section), the bulk solution concentration becomes zero after complete condensation of all counterions on macroion surface. In such a situation the exponential part of equation 3 becomes unity and thus the interactions become equivalent to non-screened Coulomb interactions. Hence the counterions on two different parts of a bent DNA can interact without screening.

As the decrease in total energy of the system with bending indicates more stability, one can refer that a charged cylinder in presence of neutralizing counterions has a propensity to bend for stability. Naturally the higher the valence the lower the energy. The drops in energy for each degree of bending are significantly large especially after the point around fr = 0.5. This picture tells that the electrostatic interaction alone is enough to cause DNA condensation especially for multivalent counterions. Figure 2 also shows that even monovalent counterions can cause the bending in strong Coulomb coupling regime provided a little uncertainty in energy profile is accepted. The little fluctuations in energy can be explained from figure 3. Even though monovalent counterions are not seen to cause condensation in relevant literatures, this study indicates that there is a faint possibility at strong Coulomb coupling regime. Actually, in this study, a complete condensation of the monovalent counterions on the macroion surface has been assumed to treat monovalent ions by the same footings as taken for multivalent counterions to examine their influence on bending of the DNA. In fact, this is only possible if $l_B$ is large enough.

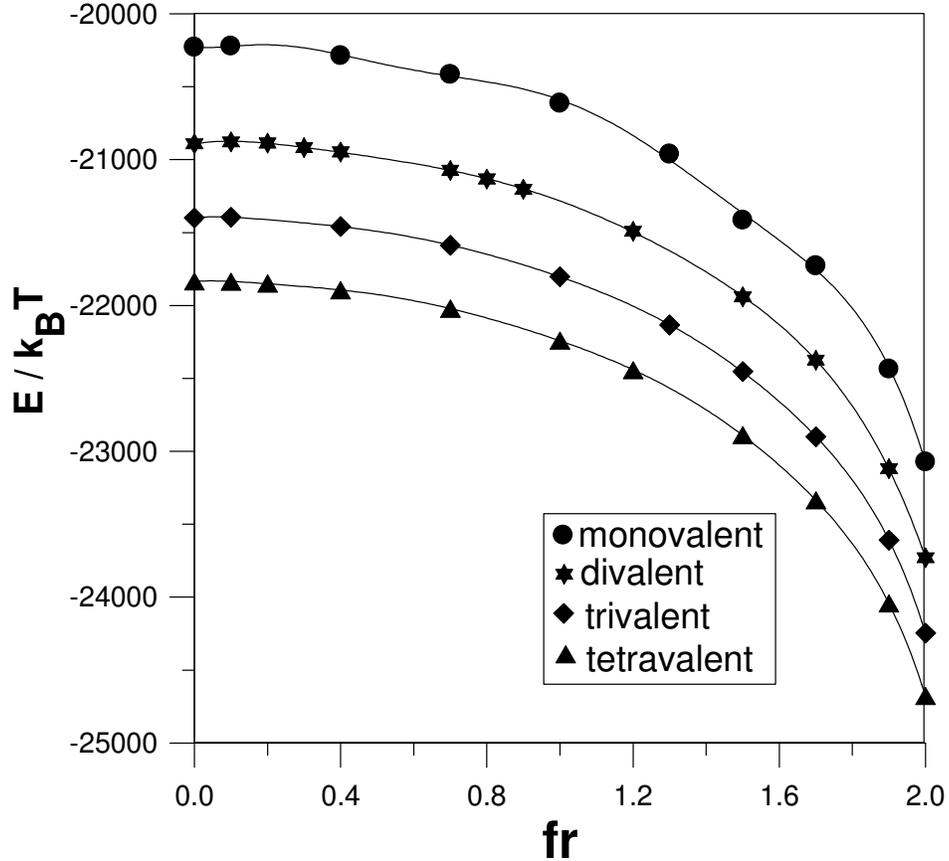

**Figure 2.** The decrease in total electrostatic potential energy (in units of $k_B T$ and Bjerrum length = 30.38 Å) with the increase of the bending fraction fr. The solid lines are polynomial fits to guide the eyes.

The energy minimized counterion positions on the charged straight cylindrical macroion (DNA) surfaces have been shown in figure 3, where usual helical patterns of distributions are observed for higher valance counterions. This picture reminds one that in case of B-DNA, cations absorbed in the grooves of the two sugar-phosphate backbones form helical patterns [11]. In the case of monovalent counterions the regular helical pattern is almost absent. This is due to their weak attractive interactions with the macroion. Another reason could be that, as stated above, all the monovalent counterions may not attached with the DNA surface as the GC length is slightly bigger than the radius of the counterions. That could be the reason behind the fluctuations in energy shown in figure 2. From these two figures one can have a conjecture that although in strong coupling regime monovalent counterions show nearly

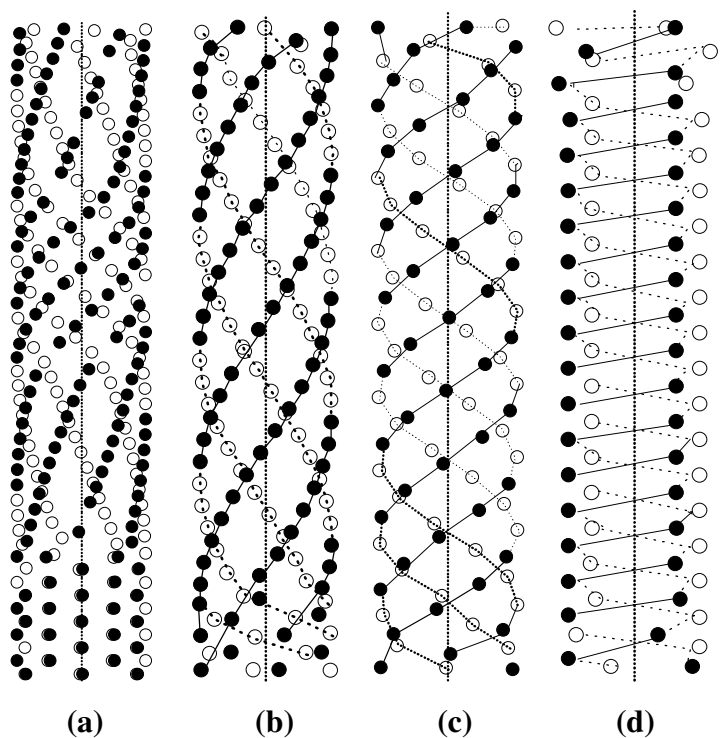

**(a)** **(b)** **(c)** **(d)**

**Figure 3.** Due to minimization of the total Electrostatic energy to the lowest possible state (nearly true ground state) counterions arrange themselves in helical patterns over a cylinder mimicking a c-DNA strand. Solid circles indicate the positions of counterions those are in front of the cylinder and the open circles indicate those are in the back side. Solid and dashed lines are drawn to show the patterns. The macroion discrete charges are shown by the dotted lines which also indicate the lengths of the macroions. The counterion distribution shrinks inward due to the end effects and the shrinking increases with valence. (a) Monovalent (b) divalent (c) trivalent (d) tetravalent counterions.

similar energy profiles as higher valence counterions, they may not cause bending (or condensation) in sufficiently weak Coulomb coupling.

Note that the average counterion distance may not be always the same (or even close) as the real distances between any two counterions arranged in helical pattern. Any counterion in helical distribution has two types of nearest neighbors – one is much shorter than the average distance while the other is bigger than that (see figure 3). Not surprisingly due to these variations the potential energy profile of a helical distribution of counterions can be very different from that of a nearly uniform distribution even though the number and valences of counterions are the same for both the distributions. Thus the general trend in theoretical analysis of estimating total energy from the average counterion counterion distance may not always be correct. To make a comparison, a nearly equidistant counterion distribution can be achieved in the following manner. First, the counterions can be distributed randomly on the macroion surface without considering macroion charge (neutral macroion) and then minimize the total repulsive energy by maximizing the distances among those. Due to mutual repulsions the counterions distribute themselves at very nearly equidistant positions (after a sufficient number of random moves). Next, one has to consider the macroion charge now interacting with the fixed (previously recorded) nearly equidistant counterions to calculate the attractive interaction. In Table 1 the repulsive, attractive and total interaction energies are shown for both helical and nearly equidistant counterion distributions over the macroion cylinder. Table 1 shows that all the interactions for the helical distribution are stronger than those of nearly equidistant distribution. One interesting point is where there is helical distribution the attractive interaction energy decreases with valance while it increases where there is nearly equidistant distribution.

**Table 1.** Electrostatic potential energies (in units of $k_B T$) of straight ($fr = 0$) DNA with counterions of different valances.

| Val | $E_{CC}^E$ | $E_{CC}$ | $E_{CM}^E$ | $E_{CM}$ | $E_{TOTAL}^E$ | $E_{TOTAL}$ |
|---|---|---|---|---|---|---|
| 1 | 16383.22 | 17200.85 | -35938.26 | -37428.77 | -19555.04 | -20227.92 |
| 2 | 15663.13 | 16573.51 | -35789.00 | -37463.40 | -20125.87 | -20889.90 |
| 3 | 15123.78 | 16090.86 | -35756.25 | -37487.40 | -20632.47 | -21397.85 |
| 4 | 14662.34 | 15674.56 | -35676.42 | -37509.27 | -21014.09 | -21834.71 |

$E_{CC}$ and $E_{CM}$ are counterion-counterion and counterion-macroion potential energies for helical counterion distributions and $E_{CC}^E$ and $E_{CM}^E$ are the same but for nearly equidistant counterion distributions. ($l_B = 30.38$ Å).

The counterion distribution patterns for trivalent counterions on carved macroions which are neither straight nor a circle ($0 < fr < 2$) are shown in figure 4 where the helical patterns are still obvious. It has been observed that for multivalent counterion distributions, there is hardly any tangible change in the usual helical distribution of the counterions due to bending from a straight cylinder up to its circular form.

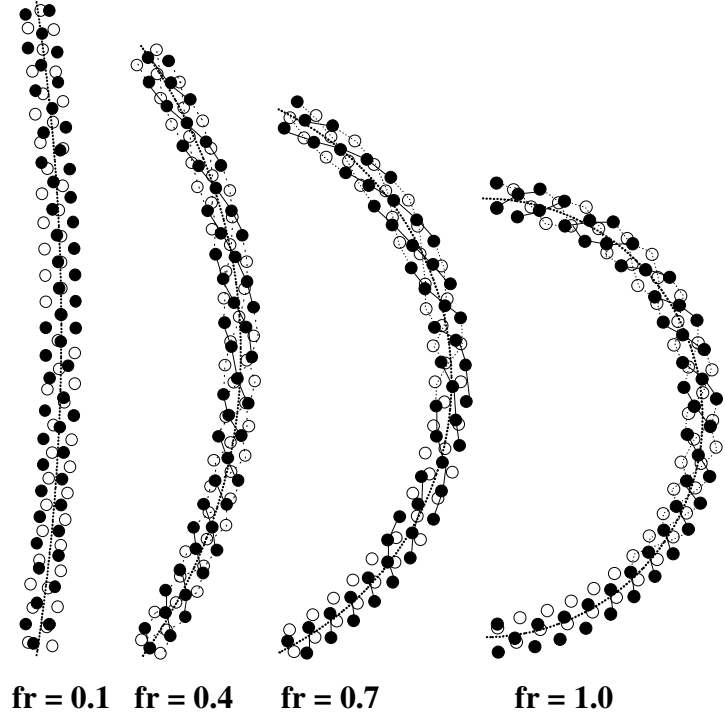

**fr = 0.1  fr = 0.4     fr = 0.7           fr = 1.0**

**Figure 4**. Trivalent counterion distributions (solid and open circles) on bent DNA. Solid circles are on the front and the open circles are on the back sides of the macroions. The macroion discrete charges are shown by dotted lines which also indicate the lengths of the macroions. Slight shrinking of counterion distributions due to end-effects is visible. Typical helical patterns of counterion distributions in each degree of bending are obvious and are very similar to those of straight DNAs. Some of the patterns have been indicated by solid and broken lines.

For curved macroions ($0 < fr \leq 2$), for a particular valance, the rate of increase of the counterion-counterion repulsive energies with bending is much slower than that of the decrease of attractive energies between the counterions and the macroion (see Table 2 and figure 5). Hence the total energy decreases with bending. As the total area over the macroion surface remains the same with bending the counterions are supposed to have essentially constant repulsive energy but the repulsion increases due to the decrease of distances between counterions of two parts of the cylinder as the

Table 2. The electrostatic attractive and repulsive potential energies (in units of $k_B T$) of a curved ($0 < fr \leq 2$) DNA with counterions of different valances.

| fr | Monovalent $E_{CC}$ | Monovalent $E_{CM}$ | Divalent $E_{CC}$ | Divalent $E_{CM}$ | Trivalent $E_{CC}$ | Trivalent $E_{CM}$ | Tetravalent $E_{CC}$ | Tetravalent $E_{CM}$ |
|---|---|---|---|---|---|---|---|---|
| 0.2 | 17217.6 | -37458.8 | 16587.8 | -37492.9 | 16103.8 | -37519.3 | 15689.8 | -37537.5 |
| 0.5 | 17295.0 | -37615.6 | 16667.4 | -37651.5 | 16182.5 | -37676.4 | 15769.7 | -37698.2 |
| 0.7 | 17390.1 | -37804.6 | 16763.4 | -37846.8 | 16278.8 | -37865.9 | 15865.7 | -37888.4 |
| 1.0 | 17605.8 | -38234.7 | 16978.5 | -38270.9 | 16495.4 | -38297.9 | 16085.6 | -38322.2 |
| 1.3 | 17935.7 | -38897.3 | 17310.5 | -38935.3 | 16828.7 | -38963.6 | 16420.7 | -38989.4 |
| 1.5 | 18251.4 | -39532.0 | 17627.4 | -39571.2 | 17145.3 | -39600.6 | 16735.4 | -39624.3 |
| 1.7 | 18685.8 | -40413.0 | 18060.9 | -40451.0 | 17572.4 | -40472.2 | 17172.6 | -40509.6 |
| 2.0 | 19912.9 | -42985.0 | 19264.0 | -42998.6 | 18770.6 | -43015.0 | 18355.9 | -43032.7 |

$E_{CC}$ and $E_{CM}$ are counterion-counterion (repulsive) and counterion-macroion (attractive) potential energies respectively ($l_B = 30.38$ Å).

direct distances get shorter than the corresponding arc lengths. This can be clearly understood from figure 4 and figure 7. The change in counterion-macroion attractive energies with valance for any degree of bending is practically negligible compared to their magnitudes (see figure 5 and Table 2). The attractive energy increases with bending due to the geometrical facility as stated below in section IV. On the other hand, counterion-counterion repulsive energy decreases with increase of valance and increases with increase of bending. The little decrease in repulsive energy with valance is due to the increase of counterion mutual distances as the counterion number decreases with increase of valance. Thus the attractive interaction is a function of bending only while the repulsive one is a function of both bending and valance.

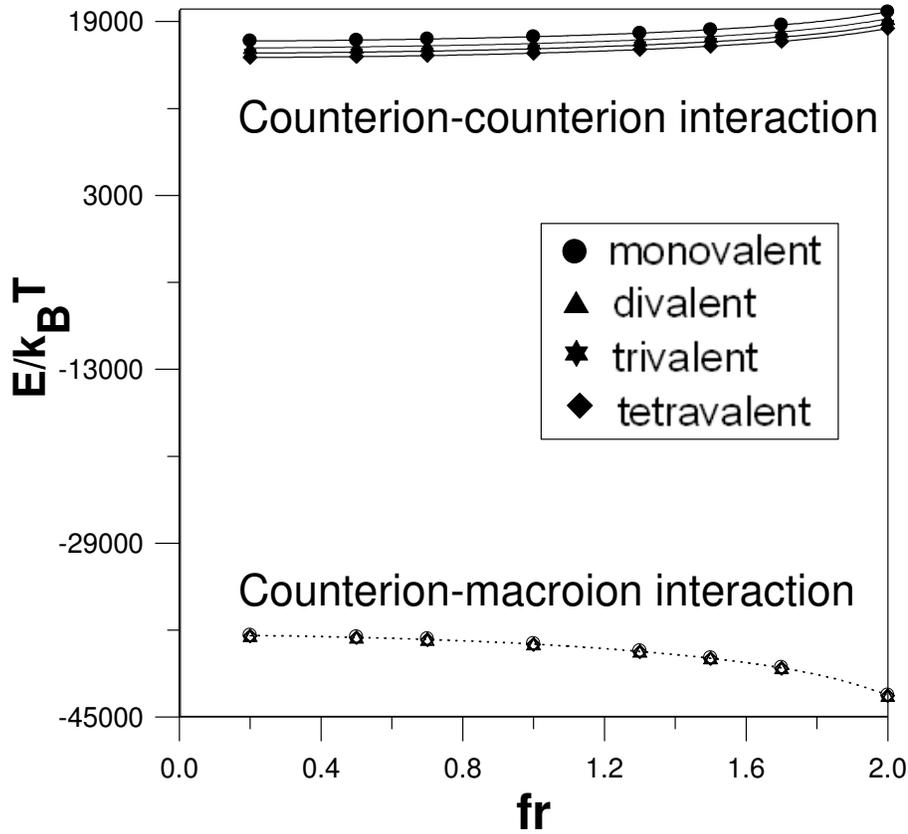

**Figure 5.** The counterion-counterion repulsive interactions (solid symbols and lines) and the counterion-macroion attractive interactions (open symbols and broken lines) for counterions of different valances versus bendings ($0 < fr < 2$). The solid and dotted lines are polynomial fits to guide the eyes. The corresponding numerical values of both types of interactions are given in Table 2.

The counterion distribution patterns on complete circular form of DNA is shown in figure 6 for all valences. Surprisingly the monovalent counterions have been observed to arrange themselves in helical pattern when the cylinder is bent to a circle. This is because of end-effects [28]. As the monovalent counterions are weakly bound with the macroion, end-effects destabilize the regular helical pattern. When the two ends meet to form the complete circle the end-effects vanish. Moreover, the macroion line charge then sets uniform circular electric field around the whole region which also helps the counterions redistribute uniformly. Due to end-effects normally the counterionion distribution shifts away from the open ends of the cylinder. Fig. 3 shows that the shifting increases with the increase in counterion valence.

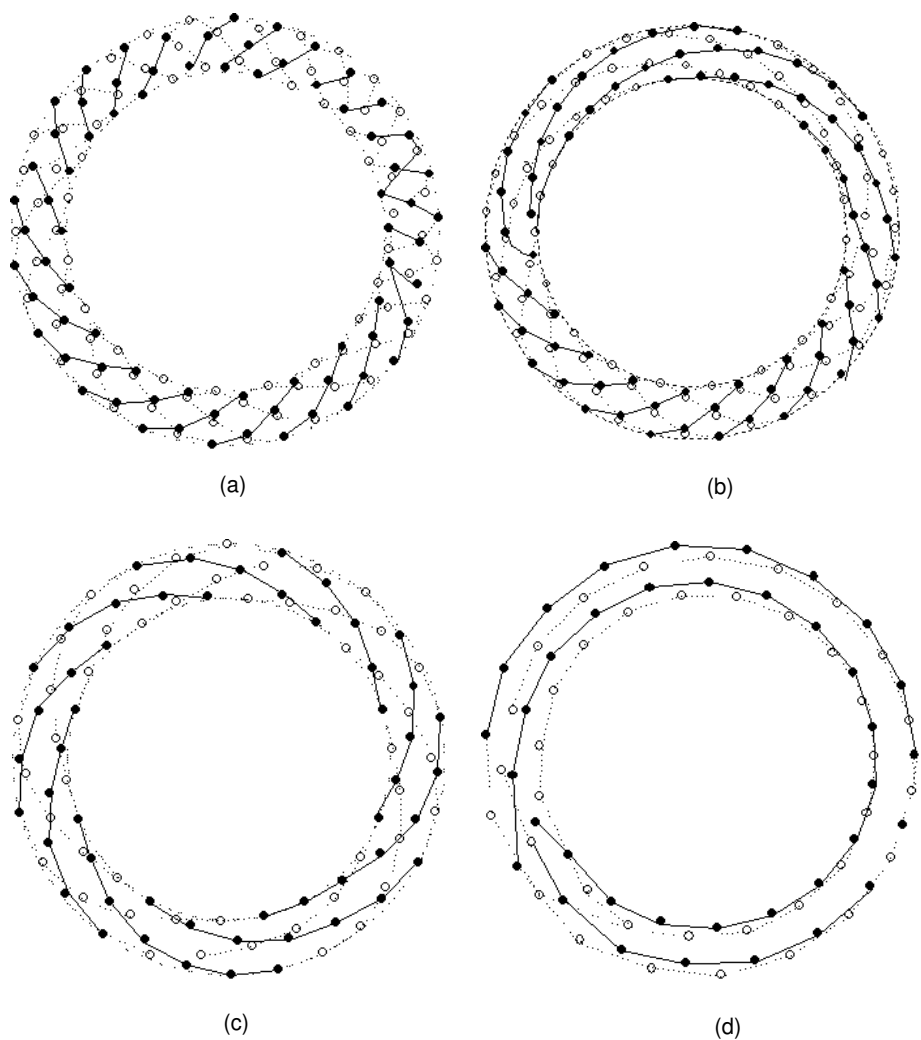

**Figure 6.** Distribution of counterions (solid and open circles) on a circular form of DNA minimizing the total electrostatic energy. (a) monovalent (b) divalent (c) trivalent (d) tetravalent counterions. Solid circles are on the front and the open circles are on the back sides of the macroions. Solid and broken lines were drawn to show the distribution patterns.

This study clearly indicates that the counterion distribution patterns are important phenomena to monitor as those are directly related to the energy profile. Since the energy minimization is nothing but the maximization of counterion mutual distances, this maximization normally yields definite counterion distribution patterns (nearly uniform, helical etc [19, 20, 31]) for regular geometries of the macroions. Thus the counterion distribution patterns not only indicate the accuracy of calculations but also determine the stability of a definite state from the detailed energy profile. It is also a known phenomenon [19] that, in strong Coulomb coupling and at nearly ground state the counterions form a glassy crystalline structure. In case of dense DNA assembly in salt solution counterion distribution patterns on DNA surfaces dictate the features of DNA-DNA electrostatic forces [17].

## IV. Geometrical View Point

As there is hardly any theory to explain the results of this study, a simple geometrical conception has been considered. As stated above, the results of the study indicate that the attractive interaction increases with bending surpluses the repulsive interaction among the counterions. One can explain it approximately by the following way:

Let's juxtapose a straight cylinder and a circular one as shown in figure 7, so that one can easily imagine the change of mutual distances between any two points on the cylinder due to bending. The length of the straight cylinder is $L = 2\pi R$.

The crucial fact behind the increase in attraction is that due to bending the distances between the counterions and the macroion charge distribution fall shorter than that of a straight cylinder. Consider that the arc length $SQ$ is equal to the distance $SQ'$ on the straight cylinder. The distance between a counterion at $P$ and a macroion point charge at $Q'$ on the straight cylinder is $\sqrt{\left(\dfrac{n}{m}L\right)^2 + r^2}$, where m is the total number of divisions on the straight cylinder and out of m, n (= m/2) is the number of divisions under consideration. This distance is constant for any other counterion, such as $P'$, on a horizontal ring around the straight cylinder containing the counterion at $P$, so the potential energy between any of those macroions and the point charge at $Q'$ are the same. For better accuracy two more counterions on the two opposite sides of the ring between $P$ and $P'$ have been considered. The distances $PQ$ and $P'Q$ are $\sqrt{R^2 + (R-r)^2 - 2R(R-r)\cos\phi}$ and $\sqrt{R^2 + (R+r)^2 - 2R(R+r)\cos\phi}$, where $\phi = \frac{n}{m}\pi.fr$. The average $(-E')$ of the potential energies between $Q$, $P$ and $Q$, $P'$ must be smaller than that $(-E)$ between $Q', P$ and $Q', P'$. For very small $\phi$, $E \cong E'$. For any certain value of fr and n, the multiplication of the ratio $\frac{E'}{E}(\geq 1)$ by the total potential energy of the straight cylinder yields the approximate total potential energy of the bent cylinder for that fr and n. The final value of the energy can be achieved by averaging over n. For better accuracy of the calculations one should consider larger values of n and m. The potential energies calculated by this way have been plotted in figure 8. For the sake of comparison the simulated potential energies (shown in figure 2) are also plotted in figure 8. The plot indicates that the calculations based on geometrical consideration are following the trend of the simulated results. However, with higher degrees of bending, it starts deviating from the simulated results and also the increase of the deviations is the same irrespective of the valances of counterions. The most probable reason behind the deviation could be the repulsion between the counterions of one part of the circular cylinder and the other, which has not been considered in the geometrical view point for simplicity. The only aim of this attempt is to show that the geometrical factors are mainly responsible for the lowering of the total energy with bending.

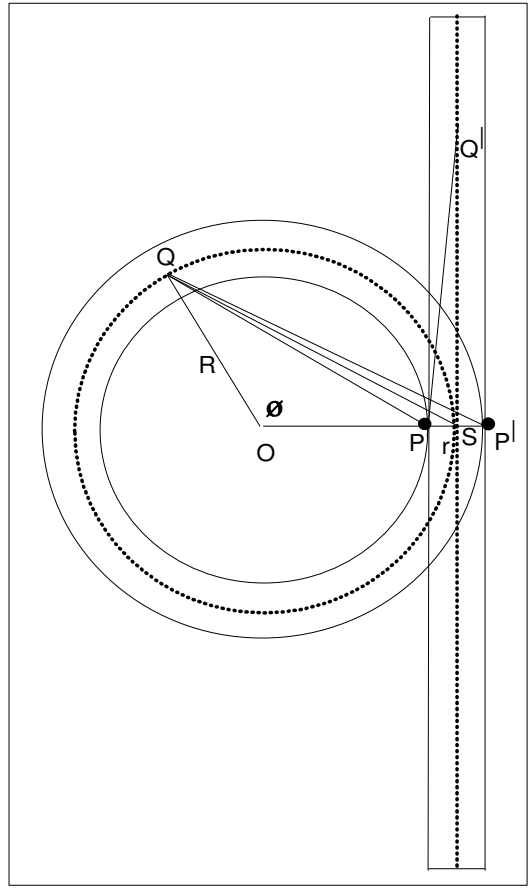

**Figure 7.** Potential energy comparison between a straight cylinder and its circular form. The dotted lines along the axis of the straight cylinder and the circular one indicate the discrete macroion charges.

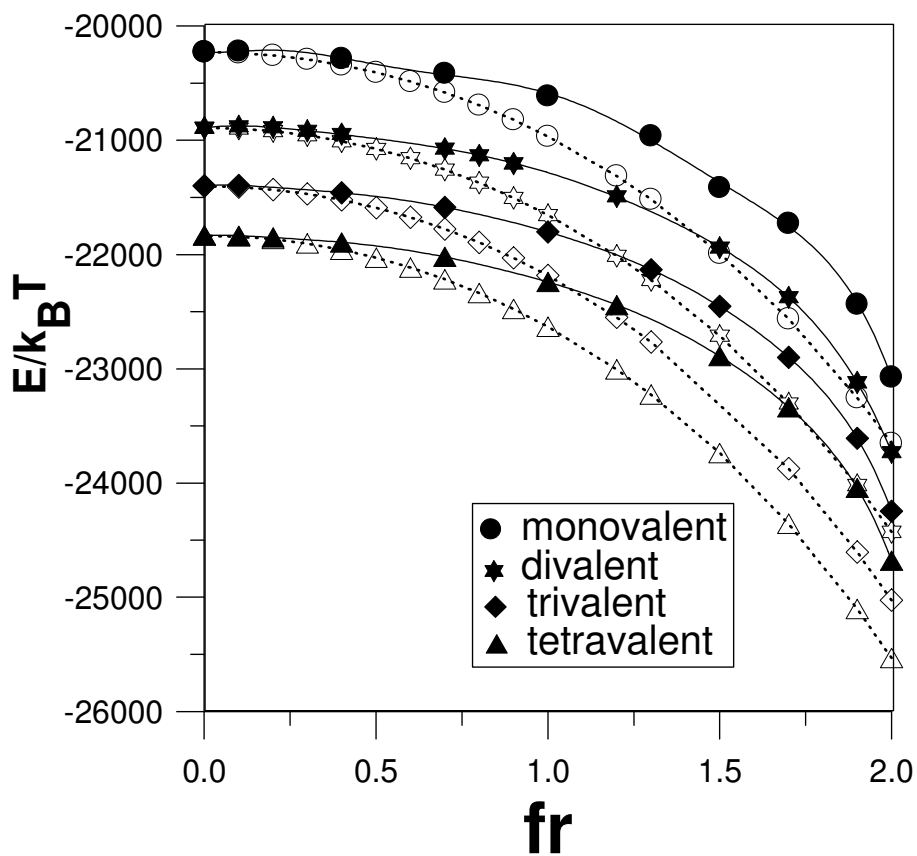

**Figure 8.** Comparison of energy (Bjerrum length = 30.34 Å) calculated from simulation (solid symbols and lines) and those calculated from geometrical view point (open symbols and dotted lines). The solid and dotted lines are polynomial fits to guide eyes.

**V. Electrostatic Ring Closure Energy**

Kunze and Netz [13] (and later employed by others [14, 15]) analytically formulated the energy difference between a straight line charge and its circular form on linearized Debye-Huckel level. Both the line and its ring shape energies are comprised of two factors, one is pure Coulomb energy and the other is exponential factor due to the screening. It is not possible to calculate the exponential factor for screening (see for example equation 3 and discussions there) directly from this study as it deals non-screened pure Coulomb interactions only. The difference of the pure Coulomb energy (per unit $k_B T$) is given by [13]

$$|\Delta E_{Coul}| = |E_{ring} - E_{rod}| = \tau^2 l_B L (1 - \ln \frac{\pi}{2}) \quad [4]$$

Where $\tau$ is the line charge density and $L$ is the length of the line charge so that $2\pi R = L$, where R is the radius of the ring. The right hand side of equation 4 is independent of valence of counterions. For $L = 510$ Å and $\tau \approx 0.59$ e/Å one can calculate $|\Delta E| \approx 96.78 l_B$. The results obtained from the data of this study are shown in Table 3 (in terms of $l_B$).

**Table 3** The simulated pure Coulomb energy factors of the electrostatic ring closure energies for different counterion valences.

| Counterions | Rod | Ring | $|\Delta E_{Coul}|$ |
|---|---|---|---|
| Monovalent | -665.83 | -759.47 | 93.64 |
| Divalent | -687.62 | -781.10 | 93.48 |
| Trivalent | -704.34 | -798.10 | 93.76 |
| Tetravalent | -718.72 | -812.35 | 93.63 |

All quantities are in terms of $l_B$.

Table 3 indicates that the simulated results are in good agreement with the predicted theoretical value (especially that the energy differences are independent of valance of the counterions). The slight deviations are almost within statistical errors of the simulation. The other probable reason of the deviation could be that the present system is a bit different from that chosen by Kunze and Netz.

## VI. Concluding Remarks

The aim of this study is, as stated in introductory section, to show that only electrostatic interactions are sufficient to condense a straight DNA and that initially a little bending (partly because of undulation or thermal agitation or electrostatic interaction itself) can cause rapid collapse to condense the DNA to a circle or even a tightly packed toroid. Apparently there can appear many other causes indicated in literatures, such as, hydration force, behind the bending but it seems that electrostatic interaction alone is enough to cause it especially in strong Coulomb coupling regime. It is important to note that the hydration force also involves electrostatics [32].

The principal achievements and findings of this study are:

A conceptually simple simulation technique [20] has been found to yield expected results related to bending of a charged cylinder (mimicking a c-DNA) by calculating the nearly ground state electrostatic potential energies in presence of neutralizing counterions in strong Coulomb coupling regime.

The energy minimized multivalent counterions on the cylindrical macroion surface have definite patterns of arrangements very similar to the helical pattern of a c-DNA and that the patterns remain almost unchanged with bending of the macroion. Similarly it can be inferred that the natural helical pattern of a c-DNA might have been due to the energy minimization for stability.

Gradual decrease in electrostatic potential energy with the decrease of radius of curvature indicates that DNA in contact with neutralizing counterions bends for stability as it is energetically fevourable. Thus the cyclization or toroidal formation of DNA in presence of counterions is an inherent characterstics. Even though it is well known that monovalent counterions do not provoke DNA condensation [33,34] but this study shows that there is a feeble possibility of the condensation by monovalent counterions in strong Coulomb coupling regime (if monovalent counterion condensation is treated in the same way as multivalent counterions).

The counterion-counterion repulsive interaction energy depends on both counterion valence and degrees of bending of the DNA while the counterion-macroion attractive interaction energy depends only on bending.

The results presented in this study considering discrete macroion charge distribution have been compared and found the same as those using continuous macroion charges for the cases of straight cylinders. It is expected that it will yield the same in cases of curved cylinders for both types of macroion charge distributions. Discrete macroion charge distribution has been considered to circumvent the problems related to integrating elliptical functions of the first kind that appears frequently for all possible counterion positions. The elliptical integrals, especially those which are not complete, i.e, solving elliptical integrals of the type $\int_0^\theta \frac{d\theta}{\sqrt{1-k^2 \sin^2 \theta}}$, when $\theta > \frac{\pi}{2}$, are often troublesome. Many researchers struggled and showed many ways to solve those types of elliptical integrals but either most of them are not simple to handle or do not match with conditions of the problem at hand. One comparatively simple way has been published recently [35] but its applicability to problems like the present study is yet to be confirmed. In future studies, the other possibility of discrete macroion charge distribution, such as, discrete charges smeared over the macroion surface [29,36] which has been reported as experimental fact [37,38,39], will be considered and compared with the results of the present study.

The comparison between the results yielded by discrete and continuous charge distributions of the straight cylinders has established that there is no need to consider the repulsive interactions among the discrete point macroion charges. It is easy to understand that had the same charge distribution was considered as continuous the question of the said repulsive interactions would not have arisen at all. Furthermore, it has been calculated and found that the total repulsive interaction among those point charges becomes enormously high and also varies significantly with the variation of the total number of point charges chosen, which is simply unphysical.

A simple and conceptual geometrical model in calculating total electrostatic energy can follow the trend of the decrease in total electrostatic energy due to bending of the straight DNA. It implies that the decrease in total energy is primarily due to the decrease in radius of curvature of the DNA and vise versa.

The above geometrical model has been considered also to check approximately the accuracy of the calculations, as it is not rational to apply theoretical approaches like mean field theories in strong Coulomb coupling regime [19,20,40-43]. In strong Coulomb coupling environment ionic correlation builds up among the counterions condensed on the macroion surface. The ionic correlation is ignored in mean field theories. Wigner crystal theory [19] is not too appropriate for narrow bent cylindrical surfaces. The Scatchard model [20] is difficult to apply to calculate the energy as it is not designed to vary with bending.

From this study it appears that helical counterion condensation on the DNA surface is the prime suspect for DNA condensation. Interestingly, it has been found before that for spherical macroion geometries the counterion distributions are nearly uniform [19]. Whereas for other macroion geometries [20,22], such as oblate spheroids, the distributions are not as uniform as that of spherical geometry. The macroion charge distribution is also an important factor that governs the counterion distribution patterns. The helical distribution has been seen only in the case where the macroion is a cylinder with a line charge distribution (continuous or discrete) along its axis.

Finally, a question related to overcharging might rise. In strong Coulomb coupling overcharging is a natural phenomenon [18,19,20] originates from ion-ion correlations. The question is will there be any spontaneous condensation in presence of multivalent counterions when the DNA is overcharged? This question will be addressed in future study but one answer can be guessed now, from the nature of overcharging, that the shape of the energy vs. fr curves (figure 2) for all valences of counterions will possibly remain the same but must will be below to those respective curves as reported in this study (e.g. the energies will be higher at each fr). As overcharging depends on counterion concentration, there might exists certain concentration of any valance and type of counterions for which DNA condensation is the most favorable.

**Acknowledgement**

AKM expresses gratitude to Dr. A. Travesset, Iowa State University and Ames Lab for providing some important journal papers and to Dr. V. A. Andreev, St. Petersburg Sate University, Russia for valuable discussions on some aspects of this study. Mr. S. Kankuzi, University of Malawi, is acknowledged for a careful reading of the manuscript.